\documentclass{ws-procs975x65}

\begin{document}

\title{In-medium QCD forces at high temperature}

\author{Yukinao Akamatsu}

\address{Kobayashi-Maskawa Institute for the Origin of Particles and the Universe (KMI), Nagoya University,\\
Nagoya 464-8602, Japan\\
$^*$E-mail: akamatsu@kmi.nagoya-u.ac.jp}

\begin{abstract}
We derive the quantum dynamics of heavy quark systems in quark-gluon plasma as open quantum systems.
The scatterings of a heavy quark with the hard medium particles give rise to Debye screened QCD force and drag force and its fluctuation.
We present a unified quantum description of these in-medium QCD forces at high temperature in the leading-order perturbation on the basis of the influence functional formalism.

\end{abstract}


\bodymatter

\section{Introduction}\label{sec:Intro}
Heavy quarks play important roles in probing properties of the quark-gluon plasma (QGP).
A single heavy quark in the QGP interacts with the medium particles and eventually it will be kinetically thermalized as is described by the Langevin equation \cite{Svetitsky:1987gq}:
\begin{eqnarray}
\frac{d\vec p}{dt}
=-\frac{\gamma}{2MT} \vec p + \vec f(t),\ \ \ 
\langle f_i(t)f_j(t')\rangle=\gamma\delta_{ij}\delta(t-t').
\end{eqnarray}
If one can observe the degree of thermalization in a given time scale, one can derive the drag force of the heavy quark.
When put in the QGP, a quarkonium, which is a bound state of a heavy quark-antiquark pair, gets unstable at sufficiently high temperature because the plasma constituents screen the color charges and turn the confining potential into the Debye screened potential \cite{Matsui:1986dk}:
\begin{eqnarray}
i\frac{\partial\Psi(r,t)}{\partial t}
=\left(-\frac{\nabla_r^2}{M}+V(r;T)\right)\Psi(r,t), \ \ \
V(r;T)=-C_{\rm F}\frac{g^2}{4\pi r}e^{-\omega_{\rm D}r}.
\end{eqnarray}
The Debye screening in the QGP and the related issues are still under active theoretical and conceptual developments 
\cite{Laine:2006ns,Brambilla:2008cx,Rothkopf:2011db,Albacete:2008dz,Akamatsu:2011se}.
If one can observe the ``melting" of quarkonium, one can infer the temperature provided that there is a reliable theory for the melting.

These heavy quark probes are fundamentally different from each other: The drag force is a classical description for the non-potential irreversible force while the melting is a quantum phenomenon of bound-states in the screened potential.
Moreover it is natural to expect that each heavy quark in a quarkonium experiences the irreversible drag process.
In a theoretical point of view, it is desirable to have a unified quantum description for both of them, whose classical limit yields the Langevin dynamics in the single heavy quark case.
In other words, we pursue the unified quantum description for in-medium forces in the QGP.
The theoretical framework needed for this purpose is that of open quantum systems.

The basics of the open quantum systems \cite{BrePetText} can be summarized as follows.
When Hilbert space consists of direct product of two Hilbert spaces ${\mathcal H}_{\rm tot}={\mathcal H_{\rm sys}}\otimes{\mathcal H_{\rm env}}$, the Hamiltonian for the total system is in general
${\bm H}_{\rm tot}={\bm H}_{\rm sys}\otimes {\bm I}_{\rm env}
+{\bm I}_{\rm sys}\otimes {\bm H}_{\rm env}+{\bm H}_{\rm int}$
and the density matrix of the total system evolves according to the von-Neumann equation:
\begin{eqnarray}
\frac{d}{dt}{\bm\rho}_{\rm tot}(t)=-i[{\bm H}_{\rm tot},{\bm\rho}_{\rm tot}(t)].
\end{eqnarray}
The dynamics of the open quantum system is described by the master equation for the reduced density matrix:
\begin{eqnarray}
\frac{d}{dt}{\bm\rho}_{\rm red}(t)={\mathcal L}{\bm\rho}_{\rm red}(t), \ \ \
{\bm\rho}_{\rm red}(t)\equiv {\rm Tr}_{\rm env}[{\bm\rho}_{\rm tot}(t)].
\end{eqnarray}
In the case of the heavy quarks in the QGP, the system degrees of freedom are the heavy quarks and the environmental degrees of freedom are the light quarks and the gluons.
In this article, we review the perturbative derivation of the master equation for the heavy quark systems at high temperature \cite{Akamatsu:2012vt}.

\section{Influence Functional of Heavy Quarks}\label{sec:IF}

The description of heavy quark systems as open quantum systems can be derived on the basis of the closed-time path formalism of non-equilibrium quantum field theory \cite{Schwinger:1960qe}.
Applied to QCD, the partition function is
\begin{eqnarray}
Z[\eta,\bar\eta]
&=&\int \mathcal {D}[\bar \psi, \psi, \bar q, q,A]
\langle \bar\psi,\bar q,A \ (s=-\infty)|{\bm\rho}_{\rm tot}|\psi, q, A \ (s=\infty)\rangle\nonumber\\
&& \times\exp
\left[i\int_{\mathcal C}ds \int d^3x\left\{
{\mathcal L}_{\rm QCD}(\bar \psi, \psi,\bar q, q, A)
+\bar \eta \psi+\bar \psi \eta
\right\}\right],
\end{eqnarray}
where $\psi$ and $q$ are the heavy and light quark Dirac fields, $A$ is the gluon field, and gauge fixing is implied in the QCD Lagrangian ${\mathcal L}_{\rm QCD}$.
Here $\mathcal C$ denotes the closed-time path contour which starts from $t=-\infty$, extends to $t=\infty$, and then returns to the original $t=-\infty$.
The initial density matrix (in the Schr\"odinger picture) is defined as ${\bm\rho}_{\rm tot}={\bm\rho}^{\rm th}_{\rm env}\otimes{\bm\rho}_{\rm sys}$: At $t=-\infty$, a system in arbitrary state ${\bm\rho}_{\rm sys}$ starts to interact with the environmental degrees of freedom in thermal equilibrium.
States $|\psi,q,A\rangle$ and $\langle\bar\psi,\bar q, A|$ are the coherent states, satisfying for example ${\bm\psi}(x)|\psi,q,A\rangle=\psi(x)|\psi,q,A\rangle$.

The dynamics of the open quantum system is described by the influence functional \cite{Feynman:1963fq} obtained after integrating out the environmental degrees of freedom.
The influence functional $S_{\rm FV}[j^{a\mu}_1, j^{a\mu}_2]$ is expressed as a functional of heavy quark color currents $j^{a\mu}_1$ on the forward time axis $\mathcal C_1$ and $j^{a\mu}_2$ on the backward time axis $\mathcal C_2$ where $\mathcal C=\mathcal C_1+\mathcal C_2$, and is expanded in terms of the currents with coefficients being the correlation functions of the gluons in thermal equilibrium.
In the leading-order perturbation and in the non-relativistic limit, $S_{\rm FV}$ (or $S_{\rm FV}^{\rm LONR}$) consists only of four-Fermi terms whose coefficients are the leading-order gluon two-point function $G^{\rm F}_{ab,00}$.
The long-time dynamics of heavy quarks is summarized in the following effective action $S_{1+2}$:
\begin{eqnarray}
S_{1+2}&\equiv& S^{\rm NR}_{\rm kin}+S^{\rm LONR}_{\rm FV}+\cdots,\\
S^{\rm NR}_{\rm kin}&=&\int d^4x\left[
\mathcal{L}^{\rm NR}_{\rm kin}(\bar\psi_1,\psi_1)-\mathcal{L}^{\rm NR}_{\rm kin}(\bar\psi_2,\psi_2)
\right],\\
S_{\rm FV}^{\rm LONR}&=&-\frac{1}{2}\int dtd^3xd^3y
(j^{a0}_{1}(t,\vec x),j^{a0}_{2}(t,\vec x))
\left(
\begin{array}{cc}
V(\vec x-\vec y) & -iD(\vec x-\vec y)\\
-iD(\vec x-\vec y) & -V^*(\vec x-\vec y) 
\end{array}
\right)
\left(
\begin{array}{c}
j^{a0}_{1}(t,\vec y) \\
j^{a0}_{2}(t,\vec y)
\end{array}
\right)\nonumber\\
&& -\int dtd^3xd^3y
\frac{\vec\nabla_x D(\vec x-\vec y)}{4T}
\cdot \left(
\vec j^{a}_{1,\rm NR}(t,\vec x)j^{a0}_2(t,\vec y)+j^{a0}_1(t,\vec x)\vec j^{a}_{2,\rm NR}(t,\vec y)
\right),
\end{eqnarray}
where $\delta_{ab}V(\vec x-\vec y)\equiv -ig^2 G^{\rm F}_{ab,00}(\omega=0,\vec x-\vec y)$, $D(\vec x-\vec y)={\rm Im}V(\vec x-\vec y)$, 
and $\vec j^{a}_{\rm NR}$ denotes the non-relativistic limit of the heavy quark color current.
With this effective action $S_{1+2}$, the partition function is given by
\begin{eqnarray}
\label{eq:HQGF}
Z[\eta_1,\bar\eta_1,\eta_2,\bar\eta_2]&=&\int \mathcal {D}[\bar \psi_1, \psi_1,\bar \psi_2, \psi_2]
\langle\bar\psi_1(-\infty)|{\bm\rho}_{\rm S}|\psi_2(-\infty)\rangle\\
&&\times \exp\Bigl[iS_{1+2}
+i\int d^4x\left(
\bar \eta_1 \psi_1+\bar \psi_1 \eta_1-\bar \eta_2 \psi_2-\bar \psi_2 \eta_2
\right)\Bigr].\nonumber
\end{eqnarray}

From Eq.~\eqref{eq:HQGF}, the time evolution of 
$\langle\psi^{\dagger}_1|{\bm\rho}_{\rm S}(t)|\psi_2\rangle$
is obtained as a ``functional Schr\"odinger  equation" with an effective Hamiltonian $\bm H_{1+2}$ derived from $S_{1+2}$.
However, the physically relevant matrix element of the density matrix is not $\langle\psi^{\dagger}_1|{\bm\rho}_{\rm S}(t)|\psi_2\rangle$ but $\rho_{\rm S}\left[t,Q^*_{1(c)},Q_{2(c)}\right]\equiv \langle Q^*_{1(c)}|{\bm\rho}_{\rm S}(t)|{Q}_{2(c)}\rangle$, where $Q_{(c)}$s are the non-relativistic Pauli spinors for the heavy (anti)quarks.
It needs to take following technical procedures to obtain the functional Schr\"odinger equation for $\rho_{\rm S}\left[t,Q^*_{1(c)},Q_{2(c)}\right]$.
In these procedures, introduction of conjugate fields $\tilde \psi_2\equiv \psi^*_2$ on $\mathcal C_2$ allows a symmetric notation between $\psi_1$ and $\tilde \psi_2$.
\begin{enumerate}
\item [(i)] 
The operator ordering in $\bm H_{1+2}$ is determined according to the time ordering of the fields in the path integral formula.
\item [(ii)] 
The functional Schr\"odinger equation for
$\rho_{\rm S}\left[t,Q^*_{1(c)},\tilde Q^*_{2(c)}\right]$
is obtained from $\bm H_{1+2}$ with substitution 
\begin{eqnarray}
(\bm Q^{\dagger}_{1(c)}, \bm Q_{1(c)}) 
\rightarrow \left(Q^*_{1(c)}, \frac{\delta}{\delta Q^*_{1(c)}}\right), 
\  
(\tilde{\bm Q}^{\dagger}_{2(c)},\tilde {\bm Q}_{2(c)})
\rightarrow \left(\tilde{Q}^*_{2(c)},-\frac{\delta}{\delta \tilde{Q}^*_{2(c)}}\right).
\end{eqnarray}
\item [(iii)] 
The above substitution gives ultraviolet divergent contributions in the four-Fermi terms arising from the Coulomb potential at the origin, e.g.
\begin{eqnarray}
&&\int d^3xd^3y V(\vec x-\vec y)\left[Q^*_{1(c)}(\vec x)t^a\frac{\delta}{\delta Q^*_{1(c)}(\vec x)}\right]\left[Q^*_{1(c)}(\vec y)t^a\frac{\delta}{\delta Q^*_{1(c)}(\vec y)}\right]\nonumber\\
&&=\int d^3x C_{\rm F}V(\vec 0) \left[Q^*_{1(c)}(\vec x)\frac{\delta}{\delta Q^*_{1(c)}(\vec x)}\right]+\cdots,
\end{eqnarray}
which needs to be renormalized.
\item [(iv)]
Instead of (ii) and (iii), it is equivalent to derive first the renormalized effective Hamiltonian $\bm H^{\rm ren}_{1+2}$, in which the operators are reordered in the normal order and the ultraviolet divergences are renormalized, and then do the substitution (ii) to obtain a functional differential operator $H^{\rm ren}_{1+2}\left[Q^*_{1(c)},\tilde Q^*_{2(c)}\right]$.
\end{enumerate}
We adopt (iv) and obtain the following renormalized effective Hamiltonian $\bm H^{\rm ren}_{1+2}$:
\begin{eqnarray}
\label{eq:ren_eff_H}
\bm H^{\rm ren}_{1+2}&=&\int d^3x
\left[
aM\left(
\bm Q_{1(c)}^{\dagger}\bm Q_{1(c)}
\right)
+\bm Q_{1(c)}^{\dagger}\left(-\frac{\nabla^2}{2M}\right)\bm Q_{1(c)}
\right]\\
&&+\int d^3x
\left[
a^*M\left(
{\tilde {\bm Q}}_{2(c)}^{\dagger} {\tilde {\bm Q}}_{2(c)}
\right)
+{\tilde {\bm Q}}_{2(c)}^{\dagger}\left(-\frac{\nabla^2}{2M}\right){\tilde {\bm Q}}_{2(c)}
\right]\nonumber\\
&&+\frac{1}{2}\int d^3xd^3y
\ {\rm N}\left\{
(\bm j^{a0}_{1}(\vec x),\bm j^{a0}_{2}(\vec x))
\left(
\begin{array}{cc}
V(\vec x-\vec y) & -iD(\vec x-\vec y)\\
-iD(\vec x-\vec y) & -V^*(\vec x-\vec y) 
\end{array}
\right)
\left(
\begin{array}{c}
\bm j^{a0}_{1}(\vec y) \\
\bm j^{a0}_{2}(\vec y)
\end{array}
\right)\right\}\nonumber\\
&&+\int d^3xd^3y
\frac{\vec\nabla_x D(\vec x-\vec y)}{4T}\cdot 
{\rm N}\left\{
\vec {\bm j}^{a}_{1,\rm NR}(\vec x)\bm j^{a0}_{2}(\vec y)
+\bm j^{a0}_{1}(\vec x)\vec {\bm j}^{a}_{2,\rm NR}(\vec y)
\right\}
+\cdots, \nonumber
\end{eqnarray}
where $a\equiv 1+C_{\rm F}V_{\rm med}/2M$, $V_{\rm med}\equiv V_{\rm}(\vec 0;T)-V_{\rm}(\vec 0;T=0)$, and $\rm N$ denotes taking the normal-ordered product.
From $\bm H^{\rm ren}_{1+2}$, the functional Schr\"odinger equation is derived as
\begin{eqnarray}
\label{eq:FSCE}
i\frac{\partial}{\partial t}\rho_{\rm S}\left[t,Q^*_{1(c)},\tilde Q^*_{2(c)}\right]
=H^{\rm ren}_{1+2}[Q^*_{1(c)},\tilde Q^*_{2(c)}]
\rho_{\rm S}\left[t,Q^*_{1(c)},\tilde Q^*_{2(c)}\right].
\end{eqnarray}
The calculation of the gluon two-point functions $G^{\rm F}_{00,ab}(\omega,\vec k)$ at leading order in the hard thermal loop resummed perturbation theory yields
\begin{eqnarray}
\label{eq:pert-v}
V(\vec r)
&=&\frac{g^2}{4\pi}
\left\{\frac{e^{-\omega_{\rm D} r}}{r}
-2iT\omega_{\rm D}^2\int^{\infty}_{0}dk
\frac{\sin kr}{r(k^2+\omega_{\rm D}^2)^2}\right\},\\
\label{eq:part-d}
D(\vec r)
&=&{\rm Im}\left[V(\vec r)\right]
=-\frac{g^2\omega_{\rm D}^2T}{2\pi}
\int^{\infty}_{0}dk\frac{\sin kr}{r(k^2+\omega_{\rm D}^2)^2},\\
\label{eq:part-a}
a&=&
1+\frac{C_{\rm F}}{2M}\frac{g^2}{4\pi}(-\omega_{\rm D}-iT), \ 
\omega^2_{\rm D}\equiv g^2T^2
\left(\frac{N_{\rm c}}{3}+\frac{N_{\rm f}}{6}\right).
\end{eqnarray}
At this order of perturbation, the effect of $t$-channel scattering between a heavy quark and hard medium particles is taken into account.
This interaction contributes to the diffusion process of the heavy quarks as well as the Debye screening of the color charges.
The former is given by the imaginary part of $V(\vec r)$ and $D(\vec r)$ while the latter is by the real part of $V(\vec r)$, and thus these in-medium forces are both contained in the renormalized effective Hamiltonian $\bm H^{\rm ren}_{1+2}$.

\section{Real-time Dynamics of Heavy Quarks}\label{sec:RTD}
The master equation for the heavy quark systems is obtained from the functional Schr\"odinger equation Eq.~\eqref{eq:FSCE}.
The matrix element of $\bm\rho_{\rm S}(t)$ between the heavy quark coherent states $\langle Q_{1(c)}^*|$ and $|\tilde Q_{2(c)}^*\rangle$ is given as a functional of the heavy quark fields by
\begin{eqnarray}
\rho_{\rm S}\left[t,Q^*_{1(c)},\tilde Q^*_{2(c)}\right]
&=&\langle Q^*_{1(c)}|\bm\rho_{\rm S}(t)| \tilde Q_{2(c)}^*\rangle,\\
\langle Q_{1(c)}^*|&\equiv&
\langle \Omega |\exp\left[-\int d^3x \left\{
\bm Q_{(c)}(\vec x)Q^*_{1(c)}(\vec x)
\right\}\right],\\
|\tilde Q_{2(c)}^*\rangle&\equiv&
\exp\left[-\int d^3x \left\{
\tilde Q_{2(c)}^*(\vec x)\bm Q_{(c)}^{\dagger}(\vec x)
\right\}\right]|\Omega\rangle.
\end{eqnarray}
Here $|\Omega\rangle$ is the heavy quark vacuum state.
By functionally differentiating $\rho_{\rm S}\left[t,Q^*_{1(c)},\tilde Q_{2(c)}^*\right]$, we obtain, for example, the reduced density matrix for a single heavy quark system:
\begin{eqnarray}
\rho^{ij}_{Q}(t,\vec x,\vec y)
&=&\langle \vec x,i|\bm \rho_Q(t)|\vec y,j\rangle
\propto \langle \Omega| \bm Q^i(\vec x) \bm\rho_{\rm S}(t)\bm Q^{j\dagger}(\vec y)|\Omega\rangle \nonumber\\
&=&-\frac{\delta}{\delta Q^{i*}_1(\vec x)}\frac{\delta}{\delta \tilde Q_2^{j*}(\vec y)}
\rho_{\rm S}\left[t,Q^*_{1(c)},\tilde Q_{2(c)}^*\right]\Big |_{Q_{1,2(c)}^*=0},
\end{eqnarray}
and that for a heavy quark-antiquark system:
\begin{eqnarray}
\rho^{ijkl}_{QQ_c}(t,\vec x_1,\vec x_2,\vec y_1,\vec y_2)
&=&\langle \vec x_1,i;\vec x_2,j|\bm \rho_Q(t)|\vec y_1,k;\vec y_2,l\rangle\\
&\propto& \langle \Omega| \bm Q^i(\vec x_1)\bm Q^j_c(\vec x_2) \bm \rho_{\rm S}(t) \bm Q_c^{l\dagger}(\vec y_2)\bm Q^{k\dagger}(\vec y_1)|\Omega\rangle \nonumber\\
&=&\frac{\delta}{\delta Q^{i*}_1(\vec x_1)}\frac{\delta}{\delta Q^{j*}_{1c}(\vec x_2)}
\frac{\delta}{\delta \tilde Q_{2c}^{l*}(\vec y_2)}\frac{\delta}{\delta \tilde Q_2^{k*}(\vec y_1)}
\rho_{\rm S}\left[t,Q^*_{1(c)},\tilde Q_{2(c)}^*\right]\Big |_{Q_{1(c)}^*=\tilde Q_{2(c)}^*=0}.\nonumber
\end{eqnarray}
The master equation for each of these reduced density matrices is therefore obtained from the functional Schr\"odinger equation for $\rho_{\rm S}\left[t,Q^*_{1(c)},\tilde Q_{2(c)}^*\right]$ in Eq.~\eqref{eq:FSCE}.

For example, let us examine the real-time dynamics of a single heavy quark system at high temperature.
The master equation for a single heavy quark system reads
\begin{eqnarray}
\label{eq:master1}
i\frac{\partial}{\partial t}\rho^{ij}_{Q}(t,\vec x,\vec y)
&=&\left\{(a-a^*)M+\left(-\frac{\nabla_x^2-\nabla_y^2}{2M}\right)\right\}\rho^{ij}_{Q}(t,\vec x,\vec y)\\
&&+\frac{1}{2}
\left(\delta_{ij}\delta_{kl}-\frac{\delta_{ik}\delta_{jl}}{N_{\rm c}}\right)
\left\{-id(\vec x-\vec y)+\frac{\vec\nabla_x d(\vec x-\vec y)}{4T}\cdot\frac{\vec \nabla_x-\vec \nabla_y}{iM}\right\}
\rho^{kl}_{Q}(t,\vec x,\vec y),\nonumber
\end{eqnarray}
and that for the color-traced reduced density matrix $\rho_{Q}(t,\vec x,\vec y)\equiv \rho^{ii}_{Q}(t,\vec x,\vec y)$ is given by
\begin{eqnarray}
\label{eq:master1_tr}
i\frac{\partial}{\partial t}\rho_{Q}(t,\vec x,\vec y)
&=&\left\{(a-a^*)M+\left(-\frac{\nabla_x^2-\nabla_y^2}{2M}\right)\right\}\rho_{Q}(t,\vec x,\vec y)\nonumber\\
&&+C_{\rm F}\left\{-id(\vec x-\vec y)+\frac{\vec\nabla_x d(\vec x-\vec y)}{4T}\cdot\frac{\vec \nabla_x-\vec \nabla_y}{iM}\right\}\rho_{Q}(t,\vec x,\vec y).
\end{eqnarray}
Interestingly, taking short distance limit $d(\vec x)\approx d(\vec 0)+d^{(2)}(\vec 0)\vec x^2/2$ in the master equation Eq.~\eqref{eq:master1_tr} reduces it to the well-known master equation of the Caldeira-Leggett model for the quantum Brownian motion \cite{Caldeira:1982iu}.
Using the master equation Eq.~\eqref{eq:master1_tr}, the following Ehrenfest relations for averaged expectation values
$\langle \bm{\mathcal{O}}\rangle(t)\equiv {\rm Tr}_x\left\{\bm\rho_{Q}(t)\bm{\mathcal{O}}\right\}=\int d^3xd^3y\langle\vec x|\bm\rho_{Q}(t)|\vec y\rangle\langle\vec y|\bm{\mathcal{O}}|\vec x\rangle$
are derived:
\begin{eqnarray}
\frac{d}{dt}\langle \vec{\bm x}\rangle&=&\frac{\langle \vec{\bm p}\rangle}{M}, \
\frac{d}{dt}\langle \vec{\bm p}\rangle=-\frac{\gamma}{2MT}\langle \vec{\bm p}\rangle,\\
\frac{d}{dt}\left\langle \frac{\bm p^2}{2M}\right\rangle
&=&-\frac{\gamma}{MT}\left(\left\langle\frac{\bm p^2}{2M}\right\rangle-\frac{3T}{2}\right).
\end{eqnarray}
To be explicit,
\begin{eqnarray}
\gamma &=&\frac{C_{\rm F}}{3}\nabla^2 d(\vec x)|_{x=0}
=-\frac{C_{\rm F}g(T)^2}{3(N_{\rm c}^2-1)}\nabla^2 \bar G^{>}_{aa,00}(\vec x)|_{x=0}\nonumber\\
&=&\frac{C_{\rm F}g(T)^2}{3(N_{\rm c}^2-1)}\int\frac{d^3k}{(2\pi)^3}k^2 G^{>}_{aa,00}(\omega=0,\vec k),
\end{eqnarray}
which is consistent with the leading-order perturbative calculation of the drag force \cite{Moore:2004tg}.
In this way, we can derive the consequences of the classical Langevin dynamics through the quantum Ehrenfest relations.

\section{Summary}\label{sec:Sum}
In this article, we reviewed the derivation of the master equation for the reduced density matrix of heavy quark systems at high temperature.
The derivation is based on the influence functional formalism applied to the heavy quark systems as the open quantum systems and is explicitly shown in the leading-order perturbation.
The essential information is contained in the renormalized effective Hamiltonian $\bm H^{\rm ren}_{1+2}$, from which the master equation is derived.

Note that this method is applicable to systems with any finite number of heavy quarks and is not limited to the leading-order perturbation; once the renormalized effective Hamiltonian is derived, it is straight forward to obtain the master equation.
As explained in Ref.~\cite{Akamatsu:2012vt}, the forward propagator of heavy quarks can also be derived from the renormalized effective Hamiltonian.
The complex potential between heavy quark-antiquark pair is defined in terms of the forward propagator and thus can also be derived from our approach to yield consistent result with Refs.~\cite{Laine:2006ns,Brambilla:2008cx} in the leading-order perturbation.
\\

Here we list several next steps of this approach:
\begin{itemize}
\item Derivation of the renormalized effective Hamiltonian in the next-to-leading order of perturbation.
We expect the next-to-leading order contribution will introduce new processes, namely the so-called gluo-dissociation processes \cite{Kharzeev:1994pz,Borghini:2011yq}.
\item Consideration of how to extend this formalism to the general non-perturbative situation.
One of the challenges is to understand how to describe heavy-light meson bound states solely in terms of heavy quark fields, or to examine what is the correct choice of the system variables as the open quantum system in such cases.
\item Phenomenological application to the dimuon spectra from bottomonia decays at the LHC \cite{Chatrchyan:2011pe}.
If one has a reliable theory of melting, one can study the physics of deconfinement by comparing with the experimental data.
\end{itemize}

\section*{Acknowledgments}
This work was supported by the Sasakawa Scientific Research Grant from the Japan Science Society.

\end{document}